 \documentclass[
twocolumn, amsmath, amssymb, floatfix,
nofootinbib,wrapfloat byrevtex]{revtex4}

 \usepackage{amsmath,mathrsfs,graphicx,epstopdf,placeins,float}
 \usepackage{booktabs}
 \usepackage{multirow}
 \usepackage{pstricks}

 \newcommand{\beq}[1]{\begin{equation}\label{#1}}
 \newcommand{\eeq}{\end{equation}}
 \newcommand{\bea}[1]{\begin{eqnarray}\label{#1}}
 \newcommand{\eea}{\end{eqnarray}}
 \newcommand{\wscr}{{\scriptstyle\mathcal{W}}}

\begin{document} 

 \title{Neural-network Quantum State of Transverse-field Ising Model}
 \author{Han-qing Shi}
 \email{hq-shi@emails.bjut.edu.cn}
\author{Xiao-yue Sun}
 \email{xy_sun@emails.bjut.edu.cn}
\author{Ding-fang Zeng\footnote{corresponding author}}
 \email{dfzeng@bjut.edu.cn}
 \affiliation{Theoretical Physics Division, College of Applied Sciences, Beijing University of Technology}

\begin{abstract}
Along the way initiated by Carleo and Troyer \cite{Science}, we construct the neural-network quantum state of transverse-field Ising model(TFIM) by an unsupervised machine learning method. Such a wave function is a map from the spin-configuration space to the complex number field determined by an array of network parameters. To get the ground state of the system, values of the network parameters are calculated by a Stochastic Reconfiguration(SR) method. We provide for this SR method an understanding from action principle and information geometry aspects. With this quantum state, we calculate key observables of the system, the energy, correlation function, correlation length, magnetic moment and susceptibility. As innovations, we provide a high efficiency method and use it to calculate entanglement entropy (EE) of the system and get results consistent with previous work very well.
\end{abstract}

\maketitle


\section{Introduction}
In a general quantum many-body system, the dimension of Hilbert space increases exponentially with the system size. W. Kohn calls this ``an Exponential Wall problem" in his Nobel Prize talks \cite{Kohn}. This lofty wall prevents physicists from extracting features and information from the system. To bypass this lofty wall, physicists make many efforts. The most productive or influential ones are density matrix renormalization group (DMRG) \cite{DMRG} and quantum monte carlo (QMC) \cite{QMC}. But till this day, no satisfactory methods are discovered for this problem universally. Each method has its advantage and disadvantages. For example, DMRG is highly efficient for 1-dimensional system, but it works not so well in higher dimensions. QMC suffers from the notorious sign problem \cite{signpro}.

However, people note that machine learning is a rather strong method for rule-drawing and information-extracting from big data sources. In this method, machine can ``learn" from data sources and ``get intelligence'', and analyze newly input data then do decisions intelligently. Very naturally, we expect machine learning may be also used to solve problems appearing in quantum many-body systems. It has been used in condensed matter physics, statistical physics, Quantum Chromodynamics, AdS/CFT, black hole physics and so on \cite{QCD,Holo,phamatt,therm,transition,blackhole}.

By our current computer power, the ``Exponential Wall problem" can not be solved through direct diagonalization of the Hamiltonian. In \cite{Laughlin}, by writing down a wave function containing enough parameters to adjust, Laughlin provides successful explanations for the fractional quantum hall effects. His doing bypass the question of exact  diagonalization of the Hamiltonian and implement a paradigm shift in the research of many-body system. After his work, people realize that the direct construction of wave function is of great value for the many-body systems' resolving. That is, we formally write down the wave function of the system that depends on enough parameters, then adjust the parameters to get the target wave function. This way, the core of the many-body problem becomes dimension reduction and feature extraction. Among the many algorithms for machine learning, artificial neural-network is a splendid one for this goal \cite{Hinton}. 

In ref.\cite{Science}, Carleo and Troyer introduced a variational representation of quantum states for typical spin models in one and two dimensions, which can be considered as a combination of Laughlin's idea and neural-networks. This neural-network quantum state (NQS) is actually a map from the spin configuration space to wave function or complex number domain. In this framework, adjusting the neural-network parameters so that for each input spin configuration, the output number is proportional to probability amplitude. In the current work, we will try out this NQS representation and machine learning method to reconstruct the ground state of the TFIM, both in one and two dimensions, and calculate its key observables, especially the EE.  For the SR method \cite{sorella1998,sorella2000,sorella2004}, we will provide an understanding basing on least action principle and information geometry.

The layout of our paper is as follows, this section is about history and motivation; the next section is a brief introduction to the neural-network quantum state and TFIM. Section III is our discussion on the SR method and its programing implementation. Section IV is our calculation results of key observables of the ground state TFIM, using machine learned NQS in section III. Section V is our method for the calculation of EE of the ground state TFIM. The last section is our summary and prospect for future work. 

\section{Neural-network quantum state and transverse-field Ising model}

The neural-network that Carleo and Troyer proposed to describe spin-$ \frac{1}{2} $ quantum system has only two layers, a visible layer  $ s=(s_1,s_2,\cdots,s_N) $ corresponding to the real system, and a hidden layer $h=(h_1,h_2,\cdots,h_M)$ corresponding to an auxiliary structure. 
 The connecting lines between the visible nodes and the hidden nodes represent interactions between them. But there are no connecting lines  inside the visible layer and hidden layer. This type of neural-network is termed as Restrict Boltzmann Machine(RBM). Its schematic diagram is shown in Fig.1. In the following, we do not distinguish between neural-network and RBM.

The many-body wave function could be understood as a map from the lattice spin configuration space to complex number field. Explicitly, this can be written as
\beq{}
\Psi(s,\wscr)=\sum_{\{h_j\}}\exp{[\sum_{i}a_i s_i+\sum_{j}b_j h_j+\sum_{i,j}w_{ij}s_i h_j]} ,
\label{nqsWfunc}
\eeq
where $s=\{s_i\}$ denotes the spin configuration and $\wscr=\{a,b,w\}$ is the weight parameters of the neural-network. Adjusting $\wscr$ is equivalent to adjusting rules of the map. And $h_{i}=\{1,-1\}$ is the hidden variables. Since there is no interactions inside the visible layer and hidden layer themselves, the summation over hidden layers spin configuration can be traced out. So the wave function can be more simply written as
\beq{}
\Psi(s,\wscr)=e^{\sum_{i}a_i s_i}\Pi_{j=1}^{M} 2\cosh[b_j+\sum_{i}w_{ij}s_i] .
\label{nqsWfuncRBM}
\eeq
Mathematically, this NQS representation can be traced back to the work of A. Kolmogorov and V. Arnold \cite{Kolmogorov,Arnold}. It is the now named Kolmogorov-Arnold representation theorem that makes the complicated higher-dimensional function's expressing as superpositions of lower-dimensional functions possible \cite{Kolmogorov1} . 

This work focus on the TFIM, whose Hamiltonian has the form
\beq{}
\mathcal{H}=-J\sum_{\langle i,j\rangle}\sigma_{i}^{z}\sigma_{j}^{z}-h\sum_{i}\sigma_{i}^{x} ,
\eeq
where $\sigma_{i}^{z}=\left(\begin{array}{cc}1&0\\0&-1\end{array}\right)$ and $\sigma_{i}^{x}=\left(\begin{array}{cc}0&1\\1&0\end{array}\right)$ are the Pauli matrixes. $J$ represent the spin coupling and $h$ represent the strengthen of the transverse field.  
For our purpose, absolute value of $J$ and $h$ do not matter, what matters is their ratio $h/J$. We will set $J=1$ and let $h$ be variables throughout this work. Interests to this model can be dated back to the 1960s work of de Gennes and others in the study of order/disorder transition in some double-well ferroelectric systems \cite{Blinc,deGennes}.  Pfeuty's work \cite{Pfeuty} in 1970 is a milestone in this area, where the one-dimensional model is solved exactly by Jordan-Wigner transformation.  His results provide us a referring standard to test the validity of our calculations. So, our interests in this work are the neural-network quantum state representation and the corresponding ML method. TFIM provides us working examples to illustrate ideas behind this method. It is believed that this new method will provide us with ways to find the new physics behind some more complicated lattice models.
 
\begin{figure}[h]
\rule{7mm}{0pt}\includegraphics[scale=0.8]{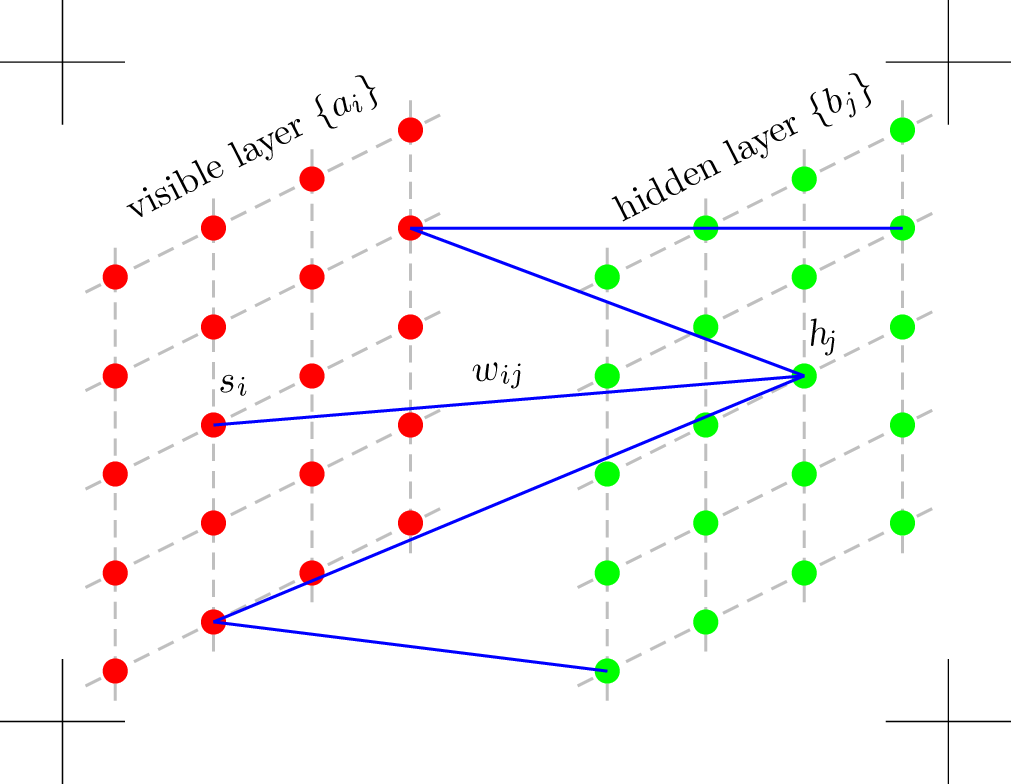}
\vspace{-15mm}
\caption{Schematic diagram of Restricted Boltzmann Machine. This is a two layer structure. The left is visible layer, the right is the auxiliary hidden layer. The dashed line between nodes in both left and right layers does not imply interactions, they are plotted here only for visual impression for `layer'. The lines between the visible nodes and hidden nodes represent interactions.}
\end{figure}

\section{Stochastic reconfiguration method for the ground state}

SR method \cite{sorella1998,sorella2000} was firstly proposed by S. Sorella and his collaborators in studies of addressing the sign problem firstly. Then it was used as an optimization method for finding goal functions from some general trial-function set. It  can be looked as a variation of the steepest descent (SD) method. Considering its key value for numerical calculation of neural-network quantum state, we provide here a new understand for it basing on  the least action principle and information geometry. Information geometry can be dated back to C. R. Rao's work in 1945 \cite{Rao}. In that work, Rao takes the Fisher information metric as Riemannian metric of statistical manifold, and regards geodesic distances as the differences between different distributions. This discipline drives to maturity after the work by Shun'ichi Amari and others \cite{Amari}. In recent years, it also gets attention as a tool to understand gravitation emergence and AdS/CFT correspondence \cite{Matsueda1,Matsueda2}. 

The quantum state of our system is functions of the neural-network parameter set $\{\wscr_{k}\}\equiv\{a_i,b_j,w_{ij}\}$. We will start from a trial function $\Psi_{T}$ which is controlled by the initial parameters $\{\wscr_k^{0}\}$. Consider a small variation of the parameters $\wscr_{k}=\wscr_{k}^{0}+\delta\wscr_{k}$, under the first order approximation, the new wave function become
\beq{}
\Psi_{T}^{'}(\wscr)=\Psi_{T}(\wscr^{0})\big[1+\sum_{k}\delta\wscr_{k}\frac{\partial}{\partial\wscr_{k}}\ln\Psi_{T}(\wscr^{0})\big] .
\eeq
Introduce a local operator $ \mathcal{O}^{k} $, so that
\beq{}
\mathcal{O}^{k}=\frac{\partial}{\partial\wscr_{k}}\ln\Psi_{T} ,
\label{Okdefinition}
\eeq
and set the identity operator $\mathcal{O}^{0}=\textbf{1}$, then $ \Psi_{T}^{'} $ can be rewritten as a more compact form
\beq{}
\Psi_{T}^{'}(\wscr)=\sum_{k}\delta\wscr_{k}\mathcal{O}^{k}\Psi_{T} .
\eeq
Our goal is to find the ground state wave function, so that the expectation value of energy $ \langle E\rangle=\frac{\langle\Psi|H|\Psi\rangle}{\langle\Psi|\Psi\rangle}$ is minimized. Obviously, $E$ depends on parameters involved in the neural-network. The procedure of looking for the ground state is equivalent to the network parameters' adjusting. The key question is the strategy of updating parameters from $\{\wscr_0\}$ to  $\{\wscr\}$. This process is something like a process that a moving from an initial point to the target point(the ground energy state in our question) in parameter space. The parameter path connecting the initial point to the target point is determined by the ``least action principle'' in parameter space as we will show as below.

In SR method, the parameters are updated by strategies
\beq{}
\wscr_{i}\longrightarrow\wscr_{i}+\Delta t\sum_{k}s_{ik}^{-1}f_{k} ,
\label{wtowNew}
\eeq
where $s_{ik}$ is the metric of the parameter space, which will be clear from the following discussion. 
Our task here is to show that this strategy is the requirement of least action principles. For this purpose, we firstly introduce generalized forces $f$
\beq{}
f_{k}=-\frac{\partial E}{\partial \wscr_{k}} .
\label{genForceDefinition}
\eeq
Then variations of the energy $E$ due to changes of $\wscr$ can be written as
\begin{align*}
\Delta E&=\frac{\Delta E}{\Delta\wscr_i}\Delta\wscr_i=\frac{\Delta E}{\Delta\wscr_i}\Delta t \cdot s^{-1}_{ik}f_k\\
&=-\Delta t \cdot s^{-1}_{ik}\frac{\Delta E}{\Delta\wscr_i}\frac{\Delta E}{\Delta\wscr_k} ,
\end{align*}
i.e.
\beq{}
\Delta E=-\Delta t\frac{(\Delta E)^{2}}{s_{ik}\Delta \wscr_i\Delta\wscr_k} .
\eeq
Now if we define $s_{ik}\Delta\wscr_i\Delta\wscr_k\equiv\Delta s$ as the line element in the parameter space, then
\beq{}
\Delta s=-\Delta E\Delta t .
\eeq
In integration form, this is nothing but,
\beq{}
\int ds=S=\int dt \mathcal{L} ,
\eeq
where $S$ is the ``action''  of the iterative process when seeking the ground state of the system and $\mathcal{L}$ is its corresponding ``Lagrangian''.  The path forms in the parameter space when the parameters are updated is determined by the corresponding least action principle. This is the physical meaning of the SR method. The SD method is a special case of SR one, whose parameter space metric is a simple Cartesian one
\beq{}
\Delta s=\sum_{k}|\wscr^{'}_{k}-\wscr_{k}|^{2} .
\eeq
However, in general cases we have no reason to take the parameter space as such a simple one. So we have to introduce a metric so that
\beq{}
\Delta s=\sum_{i,j}s_{ij}(\wscr^{'}_{i}-\wscr_{i})(\wscr^{'}_{j}-\wscr_{j}) .
\eeq
This is the reason why $s_{ik}$ appears in eq.\eqref{wtowNew}.

Obviously, $s_{ik}$'s determination is the key to the question. On this point, SR method tells us that
\beq{}
s_{ik}=\langle\mathcal{O}^{i}\mathcal{O}^{k}\rangle-
\langle\mathcal{O}^{i}\rangle\langle\mathcal{O}^{k}\rangle .
\label{sikCalcFormula}
\eeq
From information geometry's perspective, this is very natural. Consider a general data distribution $p(x;\theta)$, the Fisher information matrix or Riemannian metric on the statistic manifold is defined as. 
\begin{align}
g_{ik}(\theta)&=\int p(x;\theta)\frac{\partial \ln p(x;\theta)}{\partial \theta^{i}}\frac{\partial \ln p(x;\theta)}{\partial \theta^{k}}dx\\
&=\langle\partial_i\ln p\,\partial_k\ln p\rangle
\label{gmnInfoGeometry}
\end{align}
In our neural network quantum state, the probability reads (our wave function is limited to real fields) 
\beq{}
p(s,\wscr_{k})=\frac{|\Psi(s,\wscr_{k})|^2}{\langle\Psi|\Psi\rangle}=\frac{\Psi^2(s,\wscr_{k})}{\sum_s \Psi^2(s,\wscr_{k})}
\eeq
Substituting this into eq.\eqref{gmnInfoGeometry}, we know
\begin{align}
g_{ik}&=\sum_s p(s,\wscr)\frac{\partial\ln p(s,\wscr)}{\partial\wscr^i}\frac{\partial\ln P(s,\wscr)}{\partial \wscr^k}\\
&=\langle\mathcal{O}^{i}\mathcal{O}^{k}\rangle-\langle\mathcal{O}^{i}\rangle\langle\mathcal{O}^{k}\rangle
\nonumber
\end{align}
This is exactly the results we want to show. The rationality behind this derivation is that, mathematically a distribution function determined by its parameter set has little difference from a quantum state wave function determined by the corresponding neural-network parameters. 

Now comes our concrete implementation of the ground state finding numeric programs. The key idea is iterative execution of \eqref{wtowNew}, starting from some arbitrary point of the $\wscr\equiv\{a,b,w\}$-parameter space.  When the ground state is arrived on, the generalized force $ f=-\frac{\partial E}{\partial \wscr} $ tend to zero and the parameters are stable. Due to the exponential size of the Hilbert space, for arbitrarily chosen parameters $\wscr$, we cannot determine which state is the ground one by complete listing of all spin configurations. We use Metropolis-Hastings algorithm to sample the important configurations for approximation. The detailed step is as follows.
\begin{itemize}
\item{step~1}, starting from an arbitrary $a,b,w$ we construct $\Psi_T(s,\wscr)$ and generate $N_s=10^{3}-10^{4}$ spin state sample $\{s\}$  through a Markov chain of $s\rightarrow s'\rightarrow\cdots\rightarrow s^{(f)} $.   
The transition probability between two configurations $ s $ and $ s' $ is
\beq{}
P(s\rightarrow s')=\min(1,\lvert\frac{\Psi(s')}{\Psi(s)}\rvert^{2}) .
\eeq

\item{step~2}, for given $a,b,w$, calculate the corresponding $\mathcal{O}^{k}$,
\beq{}
\mathcal{O}^{a_i}=\frac{1}{\Psi(s)}\partial_{a_{i}}\Psi(s)=\sigma_{i}^{z} ,
\eeq
\beq{}
\mathcal{O}^{b_i}=\frac{\partial_{b_{j}}\Psi(s)}{\Psi(s)}=\tanh[b_{j}+\sum_{i}w_{ij}\sigma_{i}^{z}] ,
\eeq
\beq{}
\mathcal{O}^{w_{ij}}\!=\!\frac{\partial_{w_{ij}}\Psi(s)}{\Psi(s)}\!=\!\sigma_{i}^{z}\tanh[b_{j}+\sum_{i}w_{ij}\sigma_{i}^{z}] .
\eeq

\item{step~3}, with $\mathcal{O}^{k}$, we calculate $s_{ik}$ according to \eqref{sikCalcFormula} where $ \langle\cdots  \rangle $ means averaging over the $N_s$ samples. Get its inverse $s_{ik}^{-1}$ and update parameters $a,b,w$ through eq.\eqref{wtowNew}

\item{step~4},  repeat the above steps enough times, until the generalized force $f_{k}$ tends to zero and the parameters become iteration stable, we will get the desired parameter for ground state.
\end{itemize}
Two points are noteworthy here, 
\begin{itemize}
\item[] i) in practical calculations $ f_{k} $ takes the form of
\beq{}
f_{k}=\langle E_{loc}\rangle\langle\mathcal{O}^{k}\rangle-\langle E_{loc}\mathcal{O}^{k}\rangle .
\eeq
$ E_{loc}=\frac{\langle s|\mathcal{H}|\psi\rangle}{\psi(s)} $ is the local energy  in Variational Monte Carlo(VMC) \cite{McMillan} for each spin configuration. 

\item[] ii) using symmetries of the model to reduce the number of  parameters, which was discussed in supplementary materials of Carleo and Troyer's paper \cite{Sciencesupp}. In our models, we impose periodic boundary conditions for the lattice, so translation symmetries are used in our calculation. Due to this symmetry, the number of free components in $a_i$ is $0$, in $b_j$ is $M/N$, in $w_{ij}$ is $\alpha\times N=\dfrac{M}{N}\times N=M$, instead of $M\times N$, where $ \alpha $ is the ratio of hidden nodes number and visible nodes number.
\end{itemize}

The following is our numeric results for TFIM in both one and two dimensional square lattices. Our numerical work can be divided into three parts: 
\begin{itemize}
\item[] i) the ground state wave function training;

\item[] ii) key observables' measurement excluding EE.

\item[] iii) the EE's measurement.
\end{itemize}
In the one-dimensional model, we do the ML and measurements in three different network parameters $\alpha=1$, $2$ and $4$. Almost no superiority is observed for larger $\alpha$. For the non-entanglement-entropy observables, our results are compared with exact solutions of \cite{Pfeuty}. They coincides very well. While for the EE, we compared our results with the \cite{Cardy}, probably due to the finite size effects, our results is only qualitatively agreeing with the literature.

\section{Key observables of the TFIM ground state }

\begin{figure}[h!]
\includegraphics[scale=0.55]{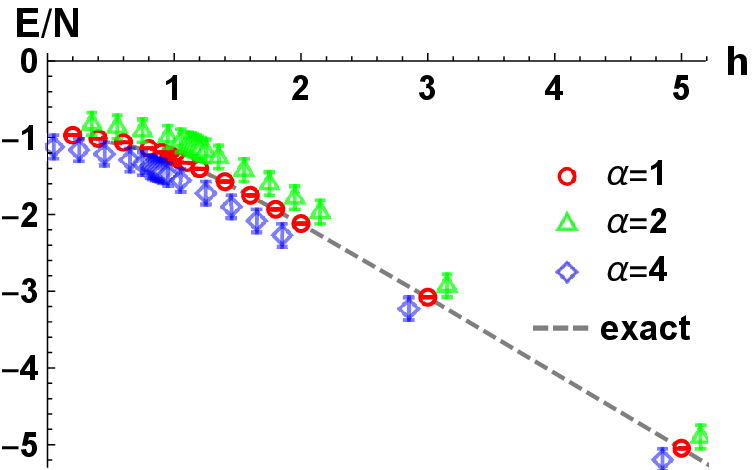}
\includegraphics[scale=0.55]{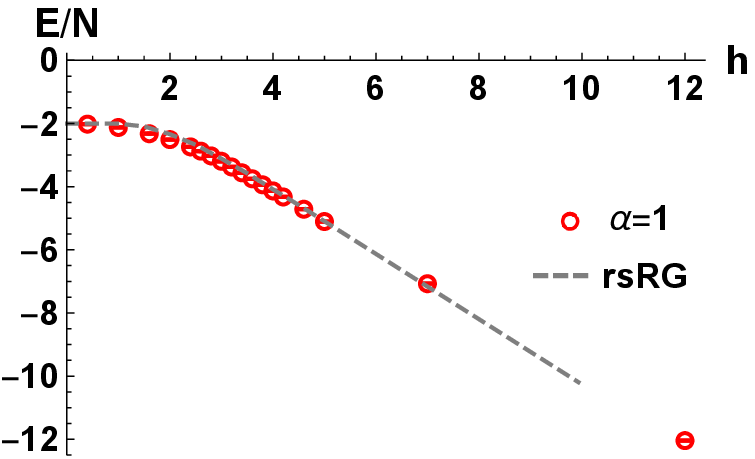}
\caption{The ground state energy $E/N$ of TFIM in one dimensional 32-site spin-chain(left) and two dimensional $10\times10$-site lattice as functions of the external field-strength. The red, green and blue points in the left figure is for network parameters $\alpha=1,2,4$. They are displaced from each other artificially otherwise coincide almost exactly. The dashed line is the analytic result of \cite{Pfeuty}. While the two dimensional result is compared with the real space renormalization group analysis of \cite{rg2dIsing1980}.}
\label{grndStateEnergy1d}
\end{figure}

Our first set of observables is the per-site ground state energy $E/N$ of TFIM for one  and two dimensional models, whose dependence on the transverse-field strength is illustrated in Fig.\ref{grndStateEnergy1d}. In one dimensional case, our numerical results fit Pfeuty's exact result \cite{Pfeuty} very good. Many numerical studies ablout 1D TFIM has been done, for example the Mote Carlo method was used in\cite{MC1dTFIM}. While in two dimension models, our results coincides with those from real space renormalization group analysis \cite{rg2dIsing1980}\cite{RGTFIM23}. In \cite{rg2dIsing1980}, the critical transverse field is 3.28. In \cite{RGTFIM23}, the critical transverse field is h=3.4351. From the figure, we easily see that the energy decreases as the field strength increases. This is because part of the energy arises from interactions between the spin sites and external magnetic fields. Very importantly, in one dimensional case we note that enlarging the network parameter $\alpha\equiv\frac{M}{N}$, i.e. the ration of hidden to manifest nodes number almost has no affects on the value and precision of the per site energy. However, the computation time grows at least linearly with $\alpha$'s growing. Due to this reason, we do not make 2-d ML and measurements for $\alpha$ greater than 1.

\begin{figure}[h!]
\includegraphics[scale=0.53]{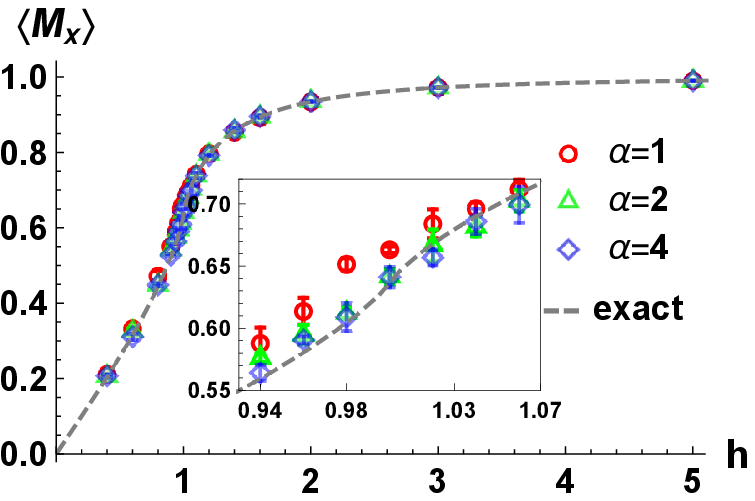}
\includegraphics[scale=0.53]{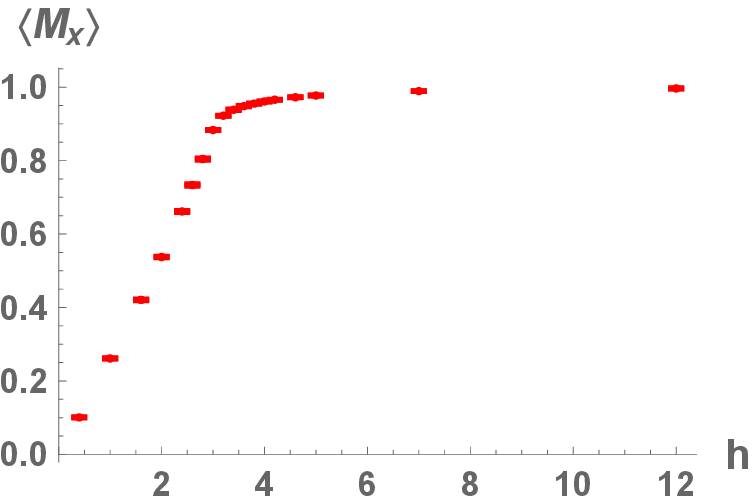}
\\
\includegraphics[scale=0.53]{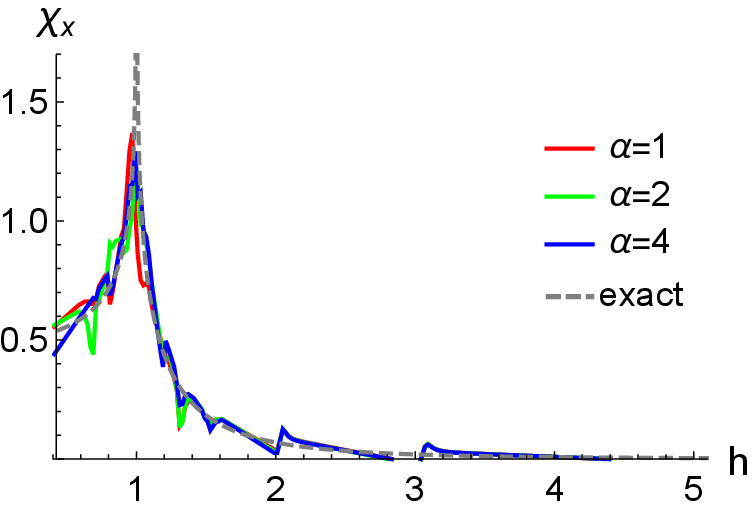}
\includegraphics[scale=0.53]{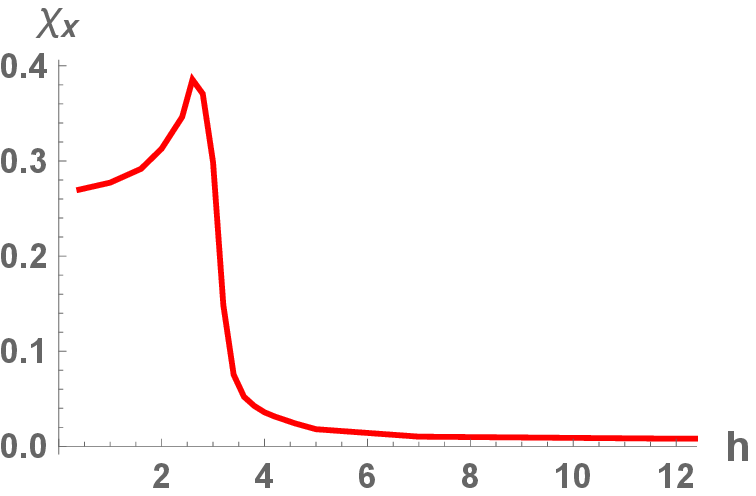}
\caption{The per-site magnet moment expectations $\langle M_{x}\rangle$ and susceptibilities $\chi_{x}$ in the one(left) and two(right) dimensional TFIM as functions of the external field-strength. The 1D model's calculation is done for three different $\alpha$'s while the 2D calculation has fixed $\alpha=1$. The dash line in the upper-left figure is the analytic results $\langle M_{x}\rangle$ from \cite{Pfeuty}. The sub figure in it shows details of conformity with the analytic result of the three $\alpha$'s numeric.}
\label{m1d}
\end{figure}

Our second set of observables is the per-site magnet moment and the corresponding susceptibility of the ground state TFIM
\beq{}
\langle M_x\rangle\!=\!\frac{\sum_i\langle\psi|\sigma^i_x|\psi\rangle}{N}
,~\chi_x\!=\!\lim_{\scriptstyle\Delta\rightarrow0}\!\!\frac{\langle M\!_x\!(h\!\!+\!\!{\scriptstyle\Delta})\!-\!M\!_x\!(h)\rangle}{{\scriptstyle\Delta}}
\eeq 
Focusing on the component along external transverse field, the results are displayed in Fig.\ref{m1d} explicitly.  For one dimensional case, our results coincide with existing literatures very well. From the figure, we see that the susceptibility contains a singularity at $h=1$(1D case) and $h\approx3$(2D case), which corresponds to the quantum phase transition points as the external field strength varies. The enlarged detail figure in this figure seems to indicate that more lager $\alpha$ ML gives $\langle M_x\rangle-h$ line more well coincides with the analytic result.

\begin{figure}[h!]
\includegraphics[scale=0.5]{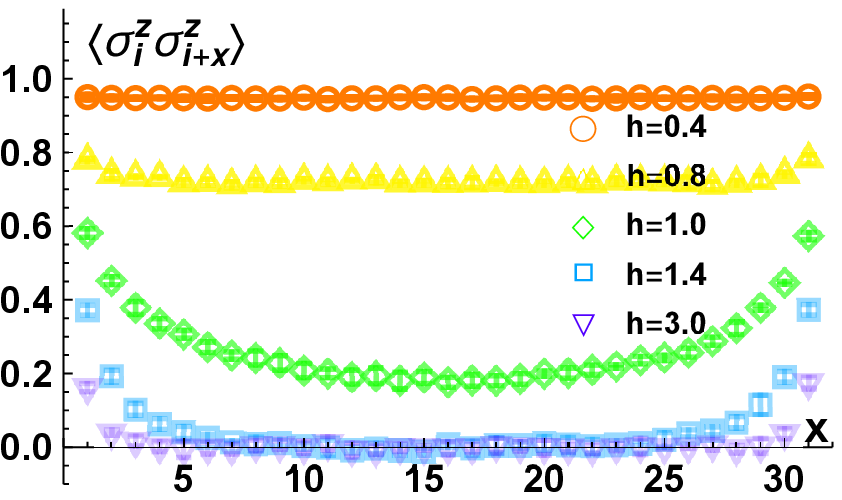}
\includegraphics[scale=0.5]{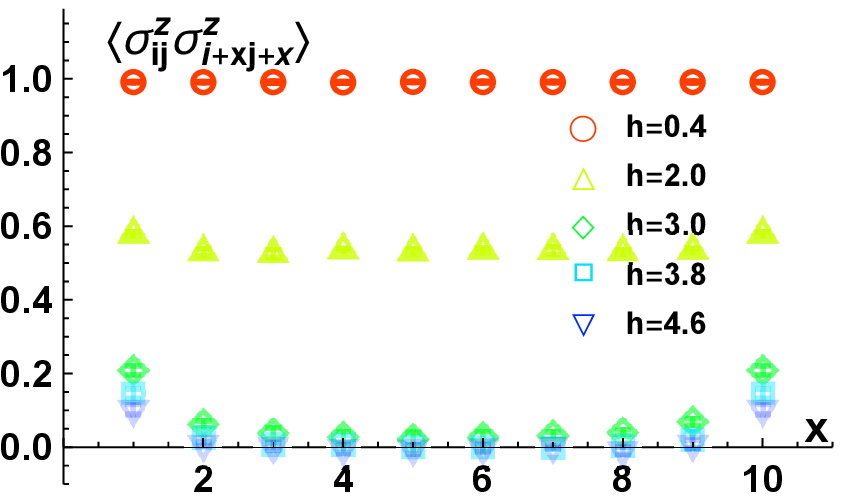}
\\
\includegraphics[scale=0.53]{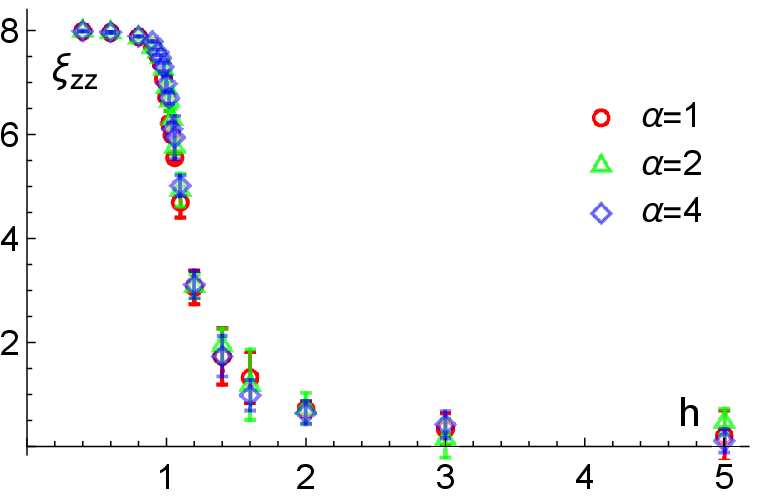}
\includegraphics[scale=0.53]{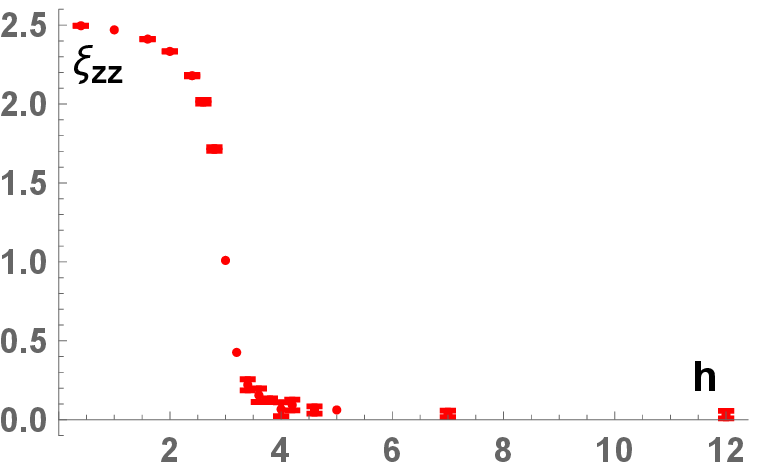}
\caption{The ground state spin-z correlation function $\langle\sigma_{i}^{z}\sigma_{i+x}^{z}\rangle$ (upper) and the corresponding correlation length (downer) of TFIM. The left is for 1D chain with 32 sites while the right is for 2D lattice of $10\times10$ sites.}
\label{cf1d}
\end{figure}

Our third set of observables is the spin-z correlation function $\langle\sigma_{i}^{z}\sigma_{j}^{z}\rangle$ and the corresponding correlation length $\xi_{zz}$, with the latter defined as $ \xi_{zz}=\frac{\langle \sum_{j}|\vec{r}_{i}-\vec{r}_{j}|\sigma_{i}^{z}\sigma_{j}^{z}\rangle}{\langle \sum_{j}\sigma_{i}^{z}\sigma_{j}^{z}\rangle} $ in numeric implementations. Our result is displayed in Fig.\ref{cf1d}. From the figure, we easily see that the system manifests long-range spin $z$-$z$ correlation in small transverse-field strength, while in large $h$ region, the correlating function decreases quickly. The correlation length $\xi_{zz}$'s behavior tells us this point more directly. In 1D case, the correlation length's jumping occurs on $h\approx1$. While in the 2D case, such jumping occurs on $h\approx3$. Due to the finite size effect of our lattice model, $\xi_{zz}$ has saturation values in the small $h$ region. This saturation phenomena will disappear in thermodynamic limits and the correlation length will diverge on the critical point.

Physically, the spin-spin interaction $J$ and the external field $h$'s influence in TFIM are two competitive factors. The former has the trends of preserving orders in the lattice, while the latter tries to break such orders. The quantum phase transition occurs on a critical value of $h/J$. In our numerics, we set $J=1$. The more larger critical value $h=3$ in 2D than the $h=1$ value in 1D is due to the more number --- twice as much --- of spin-spin interaction bonds for each site in 2D than in 1D.

\section{EE  of the ground state}
Entanglement, a fascinating and spooky quantum feature without classical correspondences, is getting more and more attentions in many physics areas \cite{Laflorencie}. People believe that many important information about a quantum system can obtained from its bipartite EE and spectrum \cite{Haldane}. More importantly, the area law feature of EE sheds new light on our understanding of holographic dualities, quantum gravity and spacetime emergence \cite{RT,Raamsdonk}. Nevertheless, we have very few ways to calculate the this quantity efficiently. Our purpose in this section is to provide a new method for its calculation in both one- and two-dimensional TFIM.

Let the total system be described by a density matrix $\rho$ and divide the system into two parts,
$ A $ and $ B $. The EE between the two is defined as follows
\beq{}
S_{A}=-tr[\rho_{A}\log \rho_{A}] ,
\eeq
where $\rho_{A}$ is the reduced density matrix of $A$, which follows from the B-part degrees of freedom's truncating. The behavior of EE is regarded as a criteria for quantum phase transition. Using conformal field theory methods, Calabrese and Cardy \cite{Cardy} calculate the EE of 1D infinitely long Ising spin chain and show that it tends to the value of $\log2$ asymptotically for $h\rightarrow0$ and tends to 0  as $h\rightarrow\infty$. In the quantum critical point $h=1$, it diverges. In \cite{Vidal} Vidal et al. show that for spin chain in the noncritical regime the EE grows monotonically as a function of the subsystem size $L$ and will get saturated at certain subsystem size $L_{0}$. At the critical value of $h$, it diverges logarithmically with singular value $L$. 

For a general state $|\Psi\rangle$ of a system consisting of two parts $A$ and $B$,
\beq{}
|\Psi\rangle=\sum_{i,j}c_{i,j}|i\rangle_{A}\otimes|j\rangle_{B} ,
\label{schmitDecomposition}
\eeq
The matrix coefficient $c_{ij}$ is the probability amplitude of a configuration whose part-A is in $i$-th spin configuration while part-B is in $j$-th spin configuration.
With the help of Singular Value Decomposition(SVD)
\beq{}
c_{ij}=U_{ik}\Sigma_{kk'}V^\dagger_{k'j} ,
\label{cij}
\eeq
where $U$, $V$ and $\Sigma$ are $d_A\times d_A $, $d_B\times d_{B}$ and $d_A\times d_B$ matrices respectively, we can diagonalize $c_{ij}$ into $\Sigma_{kk'}$ and rewrite $|\Psi\rangle$ into the form
\beq{}
|\Psi\rangle=\sum_{k=1}^{\min(d_{A},d_{B})}\sqrt{p_{k}}|\psi_{k}\rangle_{A}\otimes|\psi_{k}\rangle_{B} ,
\label{svdPsifunction}
\eeq
where $\sqrt{p_{k}}$ equals to the diagonal elements of $\Sigma_{kk'}$. Then EE between $ A $ and $ B $ can be calculated as
\beq{}
S_{A}=-\sum_{k=1}^{\min(d_{A},d_{B})}p_{k}\log p_{k} .
\label{eeCalcFormulate}
\eeq
However, the size of $c_{ij}$ increases exponentially with the number of lattices. This exponential devil makes the SVD hard to do. We hope to get a reduced coefficient matrix to approximate the original $c_{ij}$ but reserve key features of system. We want to and only can include the important elements of $c_{ij}$. Here an approximation method to bypass the exponential wall problem is needed. The exposition and schematic diagram of our idea is as follows.

Firstly, we write the general state of the lattice system with $ N $ sites as the superposition of spin configurations in descending order of $|c_{\ell}|$,
\bea{}
&&\hspace{-3mm}|\Psi\rangle=\sum_{\ell}^{2^{N}}c_{\ell}|s_{\ell}\rangle
,~
\mathrm{with}~\cdots\leqslant |c_3|\leqslant |c_2|\leqslant |c_1|
\label{configSuperposition}
\eea
Only the first $q$ configurations with the maximal $|c_\ell|$ will be produced by a Monte Carlo sampling algorithm and saved for successive computations. For the 1D TFIM with 32 lattices, $q=10^4$ is enough. $|c_{\ell}|=\psi(s_{\ell})$ here is just the value of NQS wave functions coming from MLs. When we write the subscript of $c_\ell$ as the combination of part-A's configuration-$i$ and part-B's configuration-$j$, we will get a very sparse matrix $c_{ij}\equiv c_\ell$. If we substitute this ${c}_{ij}$ into eqs.\eqref{cij}-\eqref{eeCalcFormulate}, what we get will be a very poor approximation of EE. However, if we fill the blank position of this $c_{ij}$ matrix with NQS wave function values $\psi[s_{ij\equiv\ell}]$. Our results will become much better than the those following from the original sparse matrix $c_{ij}$.

\begin{figure}[h!]
\begin{center}
\includegraphics[scale=0.53]{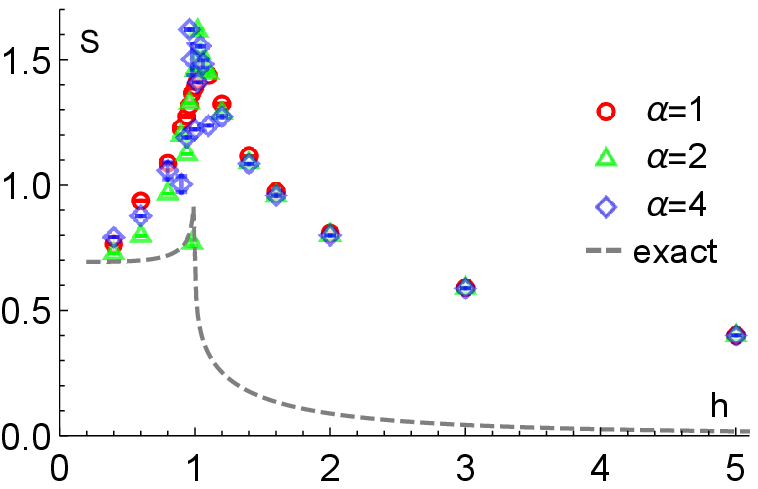}
\includegraphics[scale=0.53]{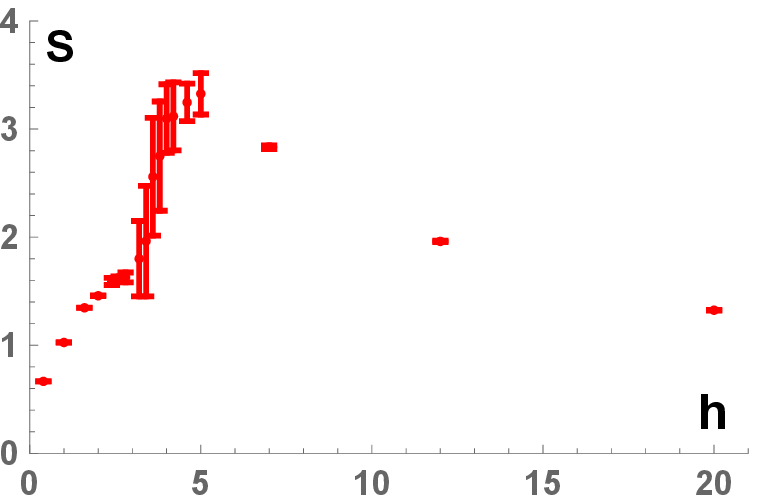}
\caption{The equal-size bipartie EE of TFIM as functions of the external field strength. The left 1D spin-chain has 32 sites and the right 2D lattice has $10\times10$ ones. In the 1D chain, three different network parameters $ \alpha=1,2,4 $ in red, green and blue is tried but the results exhibit little differences. The dashed gray line in it is the analytic result of \cite{Cardy}.}
\label{figEEh}
\end{center}
\end{figure}

Firstly, we show in FIG.\ref{figEEh} the $h$-dependence of EE when the system (both 1D chain and 2D lattice) is equally bipartite. For the 1D chain model, 3 different network parameters $\alpha=1$, $2$ and $4$ are studied and all of them yields equally good results for $S$, but the larger $\alpha$ computation costs time at least linearly increasing with $\alpha$. For this reason, we do not consider this parameters' effect on numerics for the 2D lattice. Our 1D numeric EE is compared with analytical results of \cite{Cardy,Vidal}. They have equal small-$h$ limit, approximately $S\xrightarrow{h\rightarrow0}{\log 2}$ and similar decaying trends in the large-$h$ region. They also have the same quantum phase transition point $h\approx1$. For the 2D lattice, our EE indicates that the system may experience quantum phase transitions at $h=3\sim5$. Combining with magnetic susceptibility and correlation length calculation in the previous section, we know that his transition occurs at $h\approx3$.
\begin{figure}[h!]
\begin{center}
\includegraphics[scale=0.55]{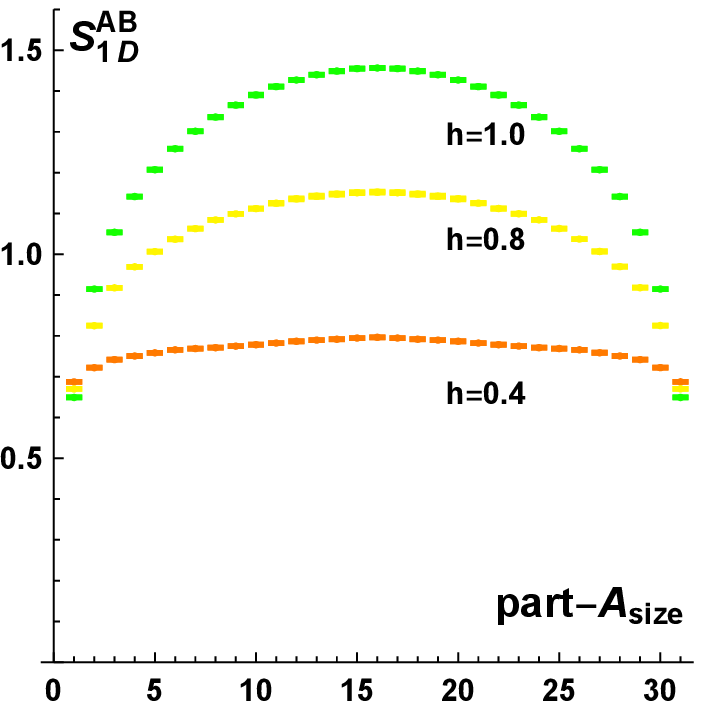}
\includegraphics[scale=0.55]{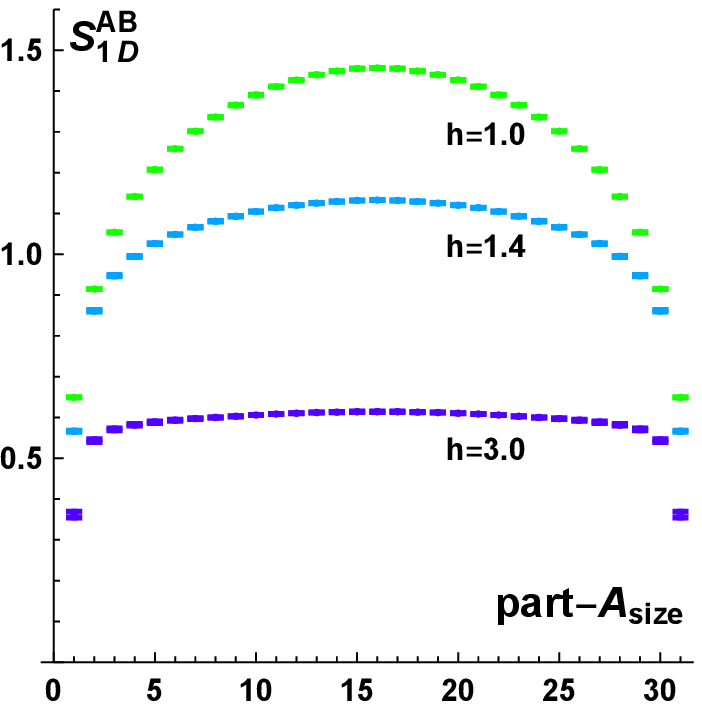}
\includegraphics[scale=0.55]{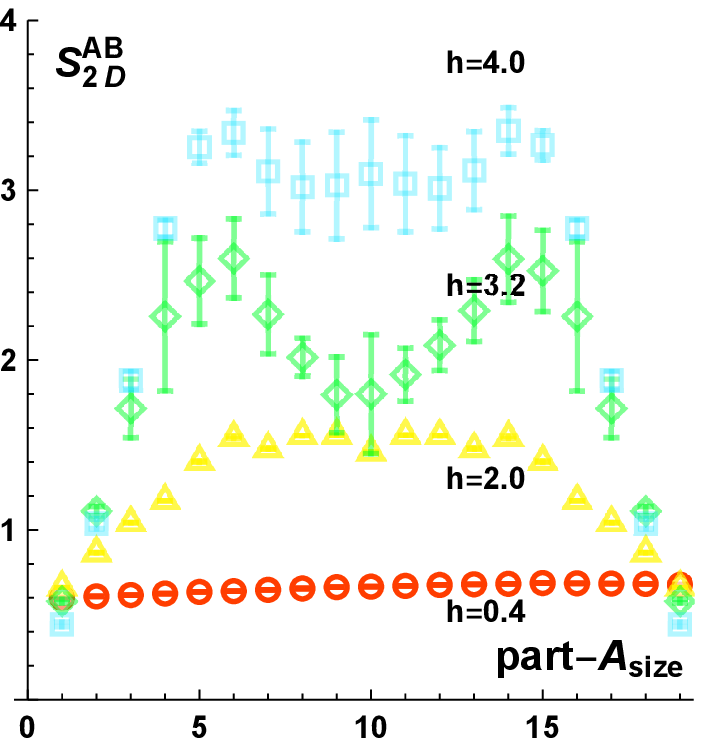}
\includegraphics[scale=0.55]{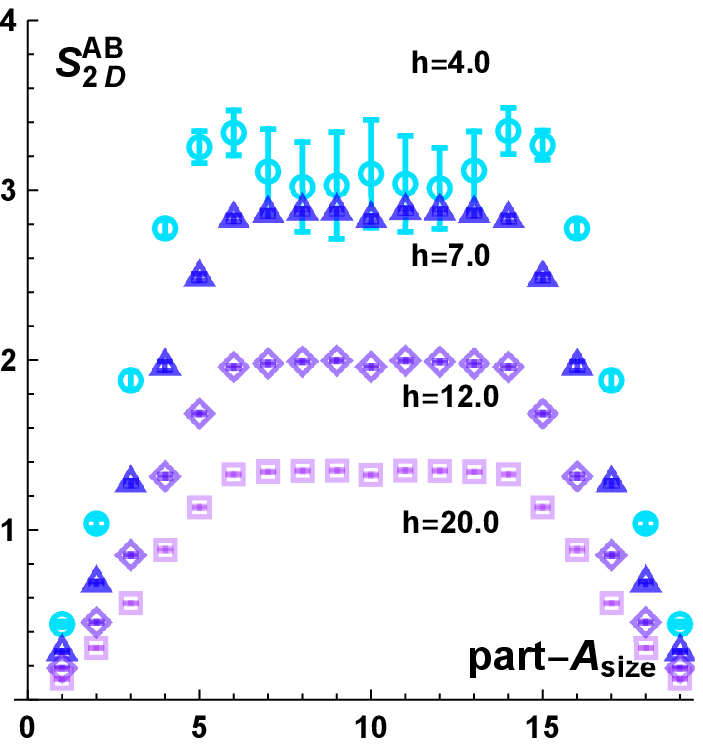}
\end{center}
\caption{The EE of TFIM as functions of the size of the A-Part in some typical transverse-field strengths. The upper is for 1D chain with 32 sites, the downer is for 2D lattice with $10\times10$ sites. In the 2D lattice, bi-partition is along the $45^\circ$ line of the lattice square.}
\label{figEEAsize}
\end{figure}

Then in FIG.\ref{figEEAsize} we studied the A-part size dependence of EE when the spin chain is arbitrarily bipartite. From the figure we see that this dependence is symmetric on the size of A and B. And, in the critical value case of the transverse-field strength, the EE increases monotonically before A-part  size increases to half the chain length. While for the much less than or more larger than the critical value of transverse-field strengths, EE rises quickly to some saturation values before the the size of A-part increases to the half size of the total system. These results agree with those of Vidal et al \cite{Vidal}.

For all known quantum many body system, their EE's calculation a challenging work. Refs.\cite{Hasting2010} and \cite{RenyiEE} are two well known works in this area. The former uses methods of QMC with an improved ratio trick sampling, whose illustrating calculation is done for 1D spin-chain of 32 sites and only the second Renyi entropy is calculated with long running times. While the latter uses wave function obtained from RBM + ML and a replica trick, also only the second Renyi entropy for a 32-sites 1D spin-chain is calculated as illustrations. As a comparison, our method can be used to calculate the von Neumann entropy for both 1D and 2D TFIM directly. Our method adopts new approximation method in the SVD approach.  We preserve the most important configurations of the system, which corresponds to the important elements of the matrix coefficient, to represent the full wave function.  The key to our method is the reduction of the matrix coefficient $c_{ij}$ and the filling of its blank positions by values of wave functions getting from RBM. In the 1D case, we get results highly agree with the known analytic results of CFT \cite{Cardy,Vidal}. While in the 2D case, our EE's calculation yields quantum phase transition signals consistent with those yielding by other observables.

\section{Conclusion and discussion}
Follow the idea of artificial neural-networks of Carleo and Troyer, we reconstruct the quantum wave function of one and two dimensional TFIM at ground state through an unsupervised ML method. Basing on the resulting wave function, we firstly calculate most of the key observables, including the ground state energy, correlation function, correlation length, magnetic moment and susceptibility of the system and get results consistent with previous works. The stochastic reconfiguration method plays key roles in the ML of neural-network quantum state representation. We provide in section II of this work an intuitive understanding for this method based on least action principle and information geometry.   As a key innovation, we provide a numeric algorithm for the calculation of EE in this framework of neural-network and ML methods. By this algorithm, we calculate entanglement entropies of the system in both one and two dimensions. 

For almost all quantum many-body system, calculations of their EE are all a challenging work to do. Both DMRG and QMC do not solve this question satisfactorily. The former works well main in 1D models, while the latter has difficulties to treat large lattice size. Our method introduced here can be used to calculate the EE directly and applies to both 1- and 2D models. On our Macbook of two core 2.9GHz CPU and 8G RAM, finishing all illustrating calculations presented in this work costs time less than five days. 

As prospects, we point here that, further exploration and revision of our numerical algorithm so that in 2D lattices it can give more clear and definite EE signals of quantum phase transition, or use our methods to study the behavior of time-dependent processes in the spin-lattice model \cite{qnchIsingNQS2018} are all valuable working directions. On the other hand, to explore the NQS representation and their ML algorithm for other physic models, such as the more general spin-lattice and Hubbard model, is obviously interesting direction to consider. For these models, more complicated neural-network such as the deep and convolution ones may be more powerful. In ref.\cite{deep}, deep neural-networks has the potential to represent quantum many-body system more effectively. While ref.\cite{CNN} shows that the combination of convolution neural-networks with QMC works even for systems exhibiting severe sign problems.


\begin{thebibliography}{99}
\bibitem{Science}
G. Carleo and M. Troyer,
``Solving the quantum many-body problem with artificial neural networks''
{\em Science} {\bf355} (2017)  602.

\bibitem{Kohn}
W. Kohn,
``Nobel Lecture: Electronic structure of matter—wave functions and density functionals'',
{\em Rev. Mod. Phys.} {\bf71} (1999) 1253.

\bibitem{DMRG}
S.R. White,
``Density matrix formulation for quantum renormalization groups'',
{\em Phys. Rev. Lett.} {\bf69} (1992) 2863.

\bibitem{QMC}
D. Ceperley and B. Alder,
``Quantum monte carlo'',
{\em Science}   {\bf231}  (1986) 555.

\bibitem{signpro}
M. Troyer and U. Wiese,
``Computational complexity and fundamental limitations to fermionic quantum Monte Carlo simulations'', {\em Phys. Rev. Lett.} {\bf94} (2005)  170201.
\href{http://arxiv.org/abs/cond-mat/0408370}{ePrint: cond-mat/0408370}

\bibitem{QCD}
P. Shanahan,  D. Trewartha and W. Detmold,
``Machine learning action parameters in lattice quantum chromodynamics'',
{\em Phys. Rev. D} {\bf97} (2018) 094506.
\href{http://arxiv.org/abs/1801.05784}{ePrint: hep-lat/1801.05784}

\bibitem{Holo}
W.C. Gan and F.W. Shu,
``Holography as deep learning''
{\em  Int. J. Mod. Phys. D} {\bf26} (2017) 1743020.
\href{http://arxiv.org/abs/1705.05750}{ePrint: gr-qc/1705.05750}

\bibitem{phamatt}
J. Carrasquilla and R. Melko,
``Machine learning phases of matter'',
{\em Nat. Phys.}{\bf13} (2017) 431.
\href{http://arxiv.org/abs/1605.01735}{ePrint: cond-mat/1605.01735}

\bibitem{therm}
G. Torlai and R. Melko,
``Learning thermodynamics with Boltzmann machines''
{\em Phys. Rev. B} {\bf94} (2016) 165134.
\href{http://arxiv.org/abs/1606.02718}{ePrint: cond-mat/1606.02718}

\bibitem{transition}
L. Wang,
``Discovering phase transitions with unsupervised learning'',
{\em Phys. Rev. B} {\bf94} (2016) 195105.
\href{http://arxiv.org/abs/1606.00318}{ePrint: cond-mat/1606.00318}

\bibitem{blackhole}
A. Askar, et al.,
``Finding Black Holes with Black Boxes -- Using Machine Learning to Identify Globular Clusters with Black Hole Subsystems'',
\href{https://arxiv.org/abs/1811.06473}{ePrint: astro-ph/arXiv:1811.06473}
\bibitem{Laughlin}
R. Laughlin,
``Anomalous quantum Hall effect: an incompressible quantum fluid with fractionally charged excitations'', 
{\em Phys. Rev. Lett.} {\bf50}  (1983) 1395.

\bibitem{Hinton}
``Hinton G E, Salakhutdinov R R. Reducing the dimensionality of data with neural networks'',
{\em Science} {\bf313} (2006) 504.

\bibitem{sorella1998}
S. Sorella,
``Green function Monte Carlo with stochastic reconfiguration'', 
{\em Phys. Rev. Lett.} {\bf80} (1998) 4558.
\href{http://arxiv.org/abs/cond-mat/9902211}{ePrint: cond-st/9902211}

\bibitem{sorella2000}
S. Sorella and L. Capriotti,
``Green function Monte Carlo with stochastic reconfiguration: An effective remedy for the sign problem'', {\em Phys. Rev. B} {\bf61} (2000) 2599.
\href{http://arxiv.org/abs/cond-mat/9902211}{ePrint: cond-mat/9902211}

\bibitem{sorella2004}
M. Casula, C. Attaccalite, and S. Sorella,
``Correlated geminal wave function for molecules: An efficient resonating valence bond approach'',
{\em J. Chem. Phys.} {\bf121}  (2004) 7110.
\href{http://arxiv.org/abs/cond-mat/0409644}{ePrint: cond-mat/0409644}

\bibitem{Kolmogorov}
A. Kolmogorov, 
``On the representation of continuous functions of several variables by superpositions of continuous functions of a smaller number of variables'',
{\em Dokl. Akad. Nauk SSSR} {\bf108}  (1961) 179.

\bibitem{Arnold}
V. Arnold,
``On functions of three variables'', 
{\em Doklady Akademii nauk SSSR} {\bf114} (1957)  679–681.

\bibitem{Kolmogorov1}
V. K$ \mathring{u}$rkov$ \acute{a} $,
``Kolmogorov's theorem and multilayer neural networks'',
{\em Neural networks}  {\bf5} (1992) 501.

\bibitem{Blinc}
R. Blinc,
``On the isotopic effects in the ferroelectric behaviour of crystals with short hydrogen bonds'',
{\em J. Phys. Chem. Solids} {\bf13} (1960) 204.

\bibitem{deGennes}
P. de Gennes
``Collective motions of hydrogen bonds'',
{\em Solid State Commun.}  {\bf1} (1963) 132

\bibitem{Pfeuty}
P. Pfeuty,
``The one-dimensional Ising model with a transverse field'',
{\em ANNALS of Physics} {\bf57} (1970) 79.

\bibitem{MC1dTFIM}
M.J.de Oliveira, J.R.N.Chiappin,
``Monte Carlo simulation of the quantum transverse Ising model''
{\em Physica A} {\bf238} (1997) 307-316.

\bibitem{Rao}
C. Rao
``Information and accuracy attainable in the estimation of statistical parameters'', 
{\em Bull Calcutta Math Soc} {\bf37} (1945) 81

\bibitem{Amari}
S. Amari and H. Nagaoka
``Methods of information geometry'',
{\em  Amer. Math. Soc.} 2007.

\bibitem{Matsueda1}
H. Matsueda,
``Emergent General Relativity from Fisher Information Metric'',
\href{http://arxiv.org/abs/1310.1831}{ePrint: hep-th/1310.1831}

\bibitem{Matsueda2}
H. Matsueda,
``Geometry and dynamics of emergent spacetime from entanglement spectrum'',
\href{http://arxiv.org/abs/1408.5589}{ePrint: hep-th/1408.5589}

\bibitem{McMillan}
W. McMillan
``Ground state of liquid He 4'',
{\em Phys. Rev.} {\bf138} (1965) A442.

\bibitem{Sciencesupp}
G. Carleo and M. Troyer,
``Supplementary Materials for Solving the quantum many-body problem with artificial neural networks'', http://science.sciencemag.org/content/sci/suppl/2017/0\\
/355.6325.602.DC1/Carleo.SM.pdf

\bibitem{Haldane}
H. Li and F. Haldane,
``Entanglement spectrum as a generalization of entanglement entropy: Identification of topological order in non-abelian fractional quantum hall effect states'',
{\em Phys. Rev. Lett.} {\bf101} (2008) 010504.
\href{http://arxiv.org/abs/0805.0332v2}{ePrint: cond-mat/0805.0332v2}

\bibitem{Laflorencie}
N. Laflorencie,
``Quantum entanglement in condensed matter systems'',
{\em Phys. Rep.} {\bf646} (2016) 1
\href{http://arxiv.org/abs/1512.03388v3}{ePrint: cond-mat/1512.03388v3}

\bibitem{RT}
S. Ryu and T. Takayanagi,
``Holographic derivation of entanglement entropy from the anti–de sitter space/conformal field theory correspondence'',
{\em Phys. Rev. Lett.} {\bf96} (2006) 181602.
\href{http://arxiv.org/abs/hep-th/0603001}{ePrint: hep-th/0603001}

\bibitem{Raamsdonk}
M. Raamsdonk,
``Building up spacetime with quantum entanglement'',
{\em Gen.Rel.Grav.} {\bf42} (2010) 2323.
\href{http://arxiv.org/abs/1005.3035}{ePrint: hep-th/1005.3035}

\bibitem{Vidal}
G. Vidal, et al.,
``Entanglement in quantum critical phenomena'',
{\em Phys. Rev. Lett.} {\bf90} (2003) 227902.
\href{http://arxiv.org/abs/quant-ph/0211074}{ePrint: quant-ph/0211074}

\bibitem{Cardy}
P. Calabrese and J. Cardy,
``Entanglement entropy and quantum field theory'',
{\em J. Stat. Mech.} {\bf0406} (2004) P06002.
\href{http://arxiv.org/abs/hep-th/0405152}{ePrint: hep-th/0405152}

\bibitem{rg2dIsing1980}
D. Mattis, J. Gallardo,
``Renormalisation group studies of two-dimensional Ising model in a transverse magnetic field'',
{\em J. Phys. C: Solid St. Phys.} {\bf 13} (1980) 2519.

\bibitem{RGTFIM23}
R. Miyazaki, H. Nishimori, and Gerardo Ortiz.
``Real-space renormalization group for the transverse-field Ising model in two and three dimensions
'',
{\em Phys. Rev. E} {\bf83} (2011) 051103. \href{https://arxiv.org/abs/1012.4557v2} {ePrint: cond-mat/1012.4557v2}

\bibitem{Hasting2010}
M. Hastings, I. Gonzalez, A. Kallin \& R. Melko,
``Measuring Renyi Entanglement Entropy in Quantum Monte Carlo Simulations'',
{\em Phys. Rev. Lett.} {\bf104} (2010) 157201,
\href{https://doi.org/10.1103/PhysRevLett.104.157201}{https://doi.org/10.1103/PhysRevLett.104.157201}

\bibitem{RenyiEE}
G. Torlai, et al.,
``Neural-network quantum state tomography for many-body systems'',
{\em Nature Phys} {\bf14} (2018) 447.
\href{https://arxiv.org/abs/1703.05334v2}{ePrint:  cond-mat/1703.05334v2}

\bibitem{deep}
L. Gao and L. Duan,
``Efficient representation of quantum many-body states with deep neural networks'',
{\em Nature Commu} {\bf8} (2017) 662.
\href{http://arxiv.org/abs/1701.05039}{ePrint: cond-mat/1701.05039}

\bibitem{CNN}
P. Broecker, et al.,
``Machine learning quantum phases of matter beyond the fermion sign problem'',
{\em Scientific Rep.}  {\bf7} (2017) 8823.
\href{http://arxiv.org/abs/1608.07848}{ePrint: cond-mat/1608.07848}

\bibitem{qnchIsingNQS2018}
S. Czischek, M. G\"arttner, T. Gasenzer,
``Quenches near Ising quantum criticality as a challenge for artificial neural networks'',
{\em Phys. Rev.} {\bf B98} (2018) no.2, 024311,
\href{http://arxiv.org/abs/arXiv:1803.08321}{1803.08321}.

\end{thebibliography}
\end{document}